\documentclass[preprint,11pt]{elsarticle}
\usepackage[english]{babel}
\usepackage{hyperref}
\usepackage{rotfloat}
\usepackage{hyperref}
\usepackage{natbib}
\usepackage{bbm}
\usepackage{bm}
\usepackage{amsmath}
\usepackage{amssymb}
\usepackage{subcaption}
\newcommand{\argmin}[1]{\underset{#1}{\operatorname{argmin}}}
\begin{document}
\begin{frontmatter}
\title{\textbf{Data Unfolding with  Mean Integrated Square Error Optimization}}
\author{Nikolay D.\ Gagunashvili\corref{author}}
\ead{nikolay@hi.is}
\address{University of Iceland, S\ae mundargata 2, 101 Reykjavik, Iceland}
\begin{abstract}
Experimental data in Particle and Nuclear physics, Particle Astrophysics and Radiation Protection Dosimetry are obtained from experimental facilities comprising a complex array of sensors, electronics and software. Computer simulation is used to study the measurement process. Probability Density Functions (PDFs) of measured physical parameters deviate from true PDFs due to resolution, bias, and efficiency effects. Good estimates of the true PDF are necessary for testing theoretical models, comparing results from different experiments, and combining results from various research endeavors.

In the article, the histogram method is employed to estimate both the measured and true PDFs. The binning of histograms is determined using the K-means clustering algorithm. The true PDF is estimated through the maximization of the likelihood function with entropy regularization, utilizing a non-linear optimization algorithm specially designed for this purpose. The accuracy of the results is assessed using the Mean Integrated Square Error.

To determine the optimal value for the regularization parameter, a bootstrap method is applied. Additionally, a mathematical model of the measurement system is formulated using system identification methods. This approach enhances the robustness and precision of the estimation process, providing a more reliable analysis of the system's characteristics.
\end{abstract}
\begin{keyword}
cluster analysis \sep 
inverse problem \sep 
system identification  \sep
entropy regularization  
\end{keyword}
\end{frontmatter}
\vspace{2cm}
\section {Introduction}
The problem of estimating an unknown Probability Density Function (PDF)  $\phi(x)$ which can be a spectrum or  a differential cross section, is one of the main subjects of analysis in Particle physics, Nuclear physics, Particle astrophysics, and   Radiation protection dosimetry. A set of independent and 
identically distributed (IID)  random variables   $\{y_1, y_2......y_{n}\}$   are commonly used. 
The random variable is a measured parameter of an event detected  in a physical experiment

A measurement is produced by an experimental setup with a complex set of sensors, electronics, and software. 
The measured  PDF $f (y)$ differs from the true PDF  $\phi(x)$   due to
the resolution of an  experimental setup, the efficiency of recording, and other factors.
As a typical example, noise components add to the kinematic parameters of a particle after it crosses the material of 
 a detector, and even the particle might not be recorded at all because of its energy loss.
 Computer modeling is used in the  study of the measurement process. 

The data under analysis can be  represented through two sets of random variables:
\begin{enumerate}
\item The measured data sample is a collection of IID random variables:
\begin{equation}
 y_1, y_2,\ldots ,y_{n} \label{meas}
\end{equation}
with the PDF $f (y)$.
\begin{itemize}
\item The true PDF $\phi(x)$ is not known.
\item The measured PDF $f(y)$  is also unknown, however, it can be estimated.
\end{itemize}
\item The simulated data sample is a collection of IID pairs of random variables:
\begin{equation}
x_1^s, y_1^s ; \, x_2^s, y_2^s; \, \ldots \,x_k^s, y_k^s  \label{simul}   
\end{equation}
with  PDF's  $\phi^s(x)$ and  $f^s(y)$. 
\begin{itemize}
\item The generated PDF  $\phi^s(x)$ is an analog of the true PDF $\phi(x)$.  
In the common case,  $\phi^s(x) \ne \phi (x)$ and the PDF $\phi^s(x)$  can be estimated. 
\item The reconstructed PDF $f^s (x)$  is an analog of the measured PDF $f(x)$ and  the PDF   $f^s (x)$ can be estimated. 
 \end{itemize}
\end{enumerate}
The simulated sample serves the purpose of constructing a mathematical model for the measurement system. 

 It is expected that the results of data analysis will be used for the interpretation of physical phenomena, e.g. to test theoretical models, 
compare results from different experiments, and to combine results from various experiments. 

A common approach to data analysis involves tackling the inverse problem to recover the true PDF, denoted as $\phi(x)$. In High Energy Physics (HEP), this process is referred to as data unfolding. 
Unfolding methods and the corresponding software applied in High Energy Physics (HEP) are comprehensively discussed in the review paper by \cite{review}. Unfolding processes can be executed using various approximations of the true PDF, such as the histogram method, kernel method, orthogonal series method, and others, as outlined in references \cite{silver} and \cite{fridman}.

Unfolding in HEP often employs the histogram method to estimate PDFs.
 The most popular and easily interpretable unfolding methods, as discussed in \cite{review}, include the Richardson-Lucy-Tarasko method \cite{taras, Richardson, Lucy}, Singular Value Decomposition method \cite{kart}, Tikhonov regularization method  \cite{Tikhonov77}, and Bin-by-Bin correction factor method  \cite{bintobin}. The expectation typically revolves around the assumption that the measurement process introduces no non-linear distortions to the measured distribution.

A commonly employed model for this case is the Fredholm integral equation, which captures the relationship between the measured PDF denoted as $f(y)$ and the true PDF  $ \phi(x)$.
\begin{equation}
\int \limits_{-\infty}^{+\infty} R(x,y)A(x) \phi(x) dx=f(y),  \,\,    \label{fred}
\end {equation}
where $A(x)$ is the probability of recording of an event with a characteristic $x$
(the acceptance); $R(x,y)$, is the probability of obtaining $y$ instead of $x$ (the
experimental resolution).
Note that for the case where $R(x,y)=R(y-x)$, a formal solution can be expressed using Fourier transformation as follows:
\begin{equation}
  \phi(x) = \frac{1}{A(x)} \int\limits_{-\infty}^{+\infty} e^{ipx}\frac{\widetilde{f}(p) }{ \widetilde{R}(p)}dp, \label{fur}
\end {equation}
where the sign $\sim$ denotes the Fourier transform of the corresponding functions. 
Indeed, the statement that the Fourier transforms of the kernel $\widetilde{R}(p)$ and the measured PDF  $\widetilde{f}(p)$ are both equal to zero within a bounded spectrum of the kernel typically implies a suppression of information in the high-frequency part of the spectra.  In essence, this implies that the measured PDF $f(y)$ lacks information about the behavior of $\phi(x)$ in the high-frequency domain. Formally, it is not feasible to deduce $\phi(x)$ from the equation (\ref{fur}) for this domain, just as it is impossible to define $\phi(x)$ for the domain where the acceptance $A(x)$ equals 0. This restriction emphasizes the limited information available in the high-frequency domain, analogous to the areas where the acceptance is zero.
{\textbf {The experimental setup has an aperture in the measurement domain and an aperture in the frequency domain.}} 

The linear  dependence of the measured PDF $f(y)$ on the true PDF $\phi(x)$  is reasonable, nevertheless in many cases 
application of the Fredholm integral  equation (\ref{fred})  is not obvious.

Let us define, in the histogram approach,  the binning $b_1,b_2,...b_{l+1}$ for the measured PDF  $f(y)$ and the binning $a_1, a_2,....a_{m+1}$  for the true PDF  $\phi(x)$.
Dependence of the measured distribution $f(y)$ from the true distribution $\phi(x)$  is approximated by a system of linear equations:
\begin{equation}
(f_1, f_2,...., f_l)^T= \mathbf {R } \times ( \phi_1,  \phi_2,..., \phi_m)^T,
\end{equation}
where  $f_i= \int\limits_{b_i}^{b_{i+1}} f(y)dy$,  $\phi_i= \int\limits_{a_i}^{a_{i+1}} \phi(x)dx$  and $\mathbf {R }$ is the response matrix, or
the linear approximation of the mathematical model of the measurement system. 
 
One crucial aspect of addressing the unfolding problem lies in the careful selection of binning, a factor that holds particular significance in multidimensional scenarios and when unfolding data from multiple detectors measuring the same spectra. 
In Section 2, we summarize the approach we will take to discretization utilizing K-mean clustering \cite{cluster}.
In Section 3, we introduce a system identification \cite{ident} approach aimed at creating a robust mathematical model for the measurement system. Post-discretization, the focus shifts to tackling the count data inverse problem, which is delved into in Section 4. Here, we explore unfolding techniques with entropy regularization.
Section 5 is dedicated to the quality assessment of the unfolding method, while Section 6 focuses on the estimation of the optimal regularization parameter. The practical application of these methods is demonstrated through a numerical example in Section 7.
\vspace{1cm}
\section {Discretization}
The quality of the unfolding solution is significantly influenced by the discretization process. In this study, we employ K-means clustering \cite{cluster} for the development of binning in the unfolding procedure. K-means clustering involves the definition of centroids, with each random value assigned to the cluster whose centroid minimally distances from the given value.

This approach to measured data, replacing bins with clusters, can be used in the case of multidimensional data, 
especially when different measurements are made of the same event data.
In contrast, for measured data, the definition of bin borders for the unfolded PDF becomes crucial. Regardless of the dimension of the unfolded distribution, the clustering method remains viable. For one-dimensional PDFs, it is straightforward, while for dimensions exceeding one, the Voronoi diagram method \cite{voronoi} proves effective.

\section {Identification of a measurement system}
One significant challenge in the unfolding process involves developing a mathematical model for the measurement system. 
Conventionally, this model is constructed using simulated data  (\ref{simul}) by method discussed in  \cite{review}. 

The method assumes that the PDF $\phi^s(x)$ do not depart too far from the unknown PDF $\phi(x)$.
 This dependence on the simulated PDF $\phi^s(x)$ is particularly pronounced in the case of wide bins, as highlighted by \cite{kuus2}. 
While the issue is less acute for narrow bins, the singularity problem is exacerbated, making the unfolding process more reliant on the strength of regularization.

Furthermore, it is assumed that the measurement process does not introduce non-linear distortions, a premise generally accepted. However, the absence of developed methods to verify this hypothesis remains a notable gap.

In order to implement the scheme outlined in Section 1, it is imperative to define the matrix $\mathbf {R }$. System identification methods, as proposed by \cite{ident}, offer a viable solution to this challenge.

System identification can be conceptually defined as the methodology employed to derive a model of a dynamic system based on observed input-output data. In the context of our application, this process aims to establish a model for the transformation of the true physical distribution into the experimentally measured distribution, a representation encapsulated by the matrix $\mathbf{R}$.

To facilitate the system identification, Monte Carlo simulation  (\ref{simul}) of a given setup serves as a valuable tool for generating the necessary input-output data. In this context, control input signals play a pivotal role in the identification process. Among the array of options, the use of impulse control signals stands out as the most popular choice, as advocated by \cite{ident}.

Let us represent  an impulse signal (input) by the content of the  generated PDF for a rather narrow bin.
 Let us have $k$ different impulse inputs that can be presented as the  $k \times m $ matrix  $\mathbf{\Phi}$,  where each of $k$ rows represents an impulse in  the original   binning for the unfolded PDF. Notice that $k$ impulses  must cover the domain of definition of the unfolded PDF.

 Denote the corresponding values of the $i$th component of the reconstructed (output) vector as
$\bm{f_i^c}=(f_{i1}^c  f_{i2}^c  \cdots f_{ik}^c)'$. For each  $i$th row of the matrix
\[ \mathbf{R} = \left(\begin{array}{llcl}
 r_{11} & r_{12} & \cdots & r_{1m} \\
 \multicolumn{4}{c}\dotfill\\
 r_{i1} & r_{i2} & \cdots & r_{im} \\
 \multicolumn{4}{c}\dotfill\\
 r_{l1} & r_{l2} & \cdots & r_{lm} \\
 \end{array}\right)\, \]
 the  equation can be written
\begin{equation}
\bm{f_i^c}=\mathbf{\Phi}\bm {r_i}+\bm{\xi_i}\, 
\end{equation}
 where
 $\bm{r_i}=( r_{i1} \, r_{i2} \cdots r_{im})'\, $,  and $\bm{\xi_i}$ is a $k$-component vector of random residuals with expectation value $\mathrm{E} \bm{\xi_i}=\bm{0}$ a variance matrix $\mathbf{\Gamma_i}$.
Formally a least squares method gives an estimate for $\bm{r_i}, i=1, \dots ,l$:
\begin{equation}
\hat{\bm{r_i}}=(\mathbf{\Phi}'\mathbf{\Gamma_i}^{-1}\mathbf{\Phi})^{-1}\mathbf{\Phi}'\mathbf{\Gamma_i}^{-1}\bm{f_i^c} \, . \label{pmatr}
\end{equation}
Estimators for the whole matrix $\mathbf{R}$ are found by producing calculations defined by equation (\ref{pmatr}) for all rows.
\section {Unfolding for the count data.}
Let us denote the number of events in bins of the measured histogram as $n_1, \ldots , n_l,n_{l+1}$, where $n_{l+1}$, represents the number of events  lost due
 to the recording inefficiency. A number of events $n_i$ for each bin can be considered as a random number with Poisson distribution. 
The likelihood function for the multivariate Poisson distribution of $n_1, \ldots , n_l, n_{l+1}$ is given by:
\begin{equation}
 \mathbb{L}(q_1, q_2,\ldots ,q_m ) = \prod_{i=1}^{l+1} \frac{\mu_i^{n_i}}{n_i!} e^{- \mu_i} \;, \label {like15}
\end{equation}
where $\mu_i = \sum_{j=1}^m r_{ij} q_j$, and $q_j$ represents the expected number of events in bin $j$ of the true distribution.

If the total number of events $n=\sum_{i=1}^{l+1}n_i$  is fixed, the distribution of events in histogram bins follows a multinomial distribution. The likelihood function can be expressed as:
\begin{equation}
 \mathbb{L}( p_1, p_2, \ldots ,p_m ) = n! \prod_{i=1}^{l+1}\frac {\theta_i^{n_i}}{n_i!} \;, \label{like21}
\end{equation}
where $\theta_i = \sum_{j=1}^m r_{ij} p_j$ and $p_j$ represents the  probability of a random event belonging to  bin $j$ of the true distribution.

The estimation of true probabilities,  involves maximizing the likelihood function (\ref{like21}). This maximization process is equivalent to the minimization of function:
\begin{equation}
 L(\mathbf{p})= -\sum_ {i=1}^{l+1}  n_i \ln \sum_{j=1}^m r_{ij}p_j.  \label{like}
\end{equation}
Unfortunately, we are confronted with an ill-posed problem that proves challenging to solve without the incorporation of regularization or the utilization of some a priori information about the solution. In this study, entropy is considered as a form of a priori information and is employed for regularization.

The entropy $H(\mathbf{p})$ of a multinomial distribution, as presented in reference \cite{mateev}, is given by:
\begin{equation}
H(\mathbf{p})=-\ln(n!) - n \sum_{i=1}^{m} p_{i} \ln(p_{i}) + \sum_{i=1}^{m} \sum_{k_{i}=0}^{n} \binom{n}{k_{i}} p_{i}^{k_{i}} (1-p_{i})^{n-k_{i}} \ln(k_{i}!). \label{mat}
\end{equation}

However, this formula can be quite complex and impractical for certain applications. Therefore, an approximation of this formula:
\begin{equation}
 H(\mathbf {p})  \approx  \frac{1}{2}\ln((2{\pi} n e)^{m-1}p_1....p_m)+\frac{1}{12n}(3m-2-\sum_{j=1}^m  \frac{1}{p_j})+O(\frac{1}{n^2}). \label {reg} 
\end{equation}
 as demonstrated in reference  \cite{multi}, is used. 
This approximation simplifies the calculation of entropy while still providing a reasonable estimate, especially for practical purposes.

In the context  of the estimation of parameters for the unfolded  PDF,  $ \hat{\mathbf{p}}$ is estimated  by minimizing the objective  function (\ref{like}) with entropy regularization defined by function (\ref{reg}),  as shown in the equation below:

\begin{equation}
\hat{\mathbf{p}} = \underset{\mathbf{p}}{\text{argmin}} [L(\mathbf{p}) + \lambda H(\mathbf{p})] \label{milke}
\end{equation}

subject to the constraints:

\begin{equation}
p_i > 0 \quad \forall i, \quad \sum p_i = 1,
\end{equation}
where $\lambda > 0$ is the regularization parameter.

Notice that the distribution  (\ref{like15}) can be represented as the product of two probability distributions \cite{baker}, namely, the Poisson distribution for the number of events $n$ and the conditional multinomial distribution for the $n$ events.

If the total number of events for the measured distribution is not known, the initial approximation can be found using simulated data  (\ref{simul}). Subsequently, an iterative procedure employing a minimization function  (\ref{milke}) , as described above, can be utilized to refine the estimation.
\vspace{1cm}
\section {Quality assessment  of unfolding  method}
There are both external and internal criteria to assess the quality of the unfolding procedure. In certain cases, external criteria exist, providing a means to gauge the assessment quality of the procedure. A classic example is the deconvolution (unfolding) of a blurred image, where the sharpness of the unblurred image serves as an external criterion for evaluating the result.

In experimental physics, defining external criteria can be challenging, particularly if a measurement has not been conducted previously. In such cases, model assumptions often offer guidance. Internal criteria for evaluating the goodness of the result become crucial in situations where reference to external information is not available.

The criteria for the quality of the data unfolding procedure presented below are integral to the process. It is noteworthy that these quality criteria can serve not only in defining the optimal value of the regularization strength in any unfolding algorithm but also for the comparison of different algorithms.
\vspace{2cm}
\subsection{Accuracy}
One commonly used measure of accuracy in the estimation of a PDF  $\hat{\phi}(x)$  is the Mean Integrated Square Error (MISE) \cite{silver, fridman}, defined as:

\begin{equation}
\mathrm{MISE} =\int \limits_{-\infty}^{+\infty} \mathbf {E}[(\hat{\phi}(x)-\phi (x))^2]dx
\end{equation}
For the histogram estimation of the unfolded distribution, MISE is expressed as:
\begin{equation}
 \begin{aligned}
\mathrm{MISE} = &\mathbf{E}  \int \limits_{a_1}^{a_{m+1}}\hat{\phi }(x)^2 dx- 2\mathbf{E}  \int \limits_{a_1}^{a_{m+1}}\hat{\phi }(x) \phi(x) dx+  \int \limits_{a_1}^{a_{m+1}}\phi (x)^2dx\\
=&\mathbf{E} \sum_{i=1}^m \frac {\hat{p_i}^2} { a_{i+1}-a_{i}} 
- 2  \mathbf{E} \sum_{i=1}^m   \frac {\hat{p_i}} { a_{i+1}-a_{i}}    \int \limits_{a_i}^{a_{i+1}} \phi (x)dx + \int \limits_{a_1}^{a_{m+1}}\phi (x)^2dx \label{msem}
\end{aligned}
\end{equation}
The last term of the expression  (\ref{msem}) does not depend on the estimators of $p_i$   (it is a constant). Therefore, the  choice an optimal parameter with the minimal value of $ \mathrm{MISE}$ does not depend on this part of equation.

Another popular measure of the accuracy  \cite{review} used in unfolding with histogram estimation of the PDF   is the Mean Squared Error           ($\mathrm{MSE}$):

\begin{equation}
 \begin{aligned}
\mathrm{MSE} =    \mathbf{E}\sum_{i=1}^m  \left (  \int \limits_{a_i}^{a_{i+1}} \hat{\phi}(x)dx-   \int \limits_{a_i}^{a_{i+1}} \phi (x)dx\right )^2\\
=&\mathbf{E} \sum_{i=1}^m \left ( \frac {\hat{p_i}} { a_{i+1}-a_{i}} -   \int \limits_{a_i}^{a_{i+1}} \phi (x)dx\right )^2
 \end{aligned}
\end{equation}
The main disadvantage of MSE is that it cannot be used to compare the accuracy of two unfolded distributions obtained with different binning schemes. This capability is important for selecting a better binning scheme for an unfolding procedure.

\subsection{Pearson's goodness of fit test statistics}
The test statistic is defined as:
 \begin{equation}
X^2(\mathbf{\hat{p}}) =\sum_{i=1}^{l+1} \frac{(n_i-n\sum_{j=1}^m r_{ij}\hat{p_j})^2} {n\sum_{j=1}^m r_{ij}\hat{p_j}}.
 \end{equation}
    The test statistic $X^2$ evaluates the discrepancy between the measured data and the unfolded distribution, taking into account the response matrix and the regularization parameter $\lambda$.
    It's useful for assessing the dependence of the unfolding solution on the strength of regularization and can aid in selecting an appropriate regularization parameter.
\subsection {Pearson's residuals} 
Pearson's residuals provide insights into the quality of fitting the measured data using the unfolded distribution folded with a response matrix.
 These residuals, denoted as:
\begin{equation}
res_i=  \frac{n_i- n\sum_{j=1}^m r_{ij}\hat{p_j }}{\sqrt{n \sum_{j=1}^m r_{ij} \hat{p_j }}}  \quad\quad  i=1,...,l+1.
\end{equation}
Residuals are used to test statistical hypotheses about the quality of the fitting, particularly regarding their independence and normal distribution with a mean of 0 and a standard deviation of 1.   To assess these assumptions, the Kolmogorov-Smirnov (KS) test and Q-Q plot can be employed.
\subsection{Kullback-Leibler divergence}
Kullback-Leibler (KL) divergence for multinomial distribution \cite{nielsen}  defined as: 
\begin{equation}
S(\mathbf{\hat{p})}=n\sum_{i=1}^m\hat{p}_i \ln \frac{\hat{p}_i}{\epsilon_i},
\end{equation}
where $\epsilon_i$ represent prior distribution, contains a priori information about true distribution. 
The KL divergence provides a quantitative measure of the difference between the estimated and prior distributions, which can guide the selection of the regularization parameter.
\section {Estimation of optimal regularization parameter}
In this paper, we establish the optimal estimation of the true (PDF) as the estimation with the minimum MISE.
To find the regularization parameter that provides a close-to-optimal PDF  in the practical case when the true PDF $\phi(x)$ is unknown, the bootstrap method  \cite{boot} can be employed. 

 Resampling of the measured distribution is performed  by generating multinomial distributions  \cite{cern} with probabilities  $\hat{\theta}_i = \sum_{j=1}^m r_{ij} \hat{p}_j (\lambda_ {\textrm{bs}}) $.  Here, $\lambda_{\textrm{bs}}$ is the lowest value of the regularization parameter with $\hat{\theta}_i(\lambda_{\textrm{bs}})$ 
that satisfies the quality assessment criteria presented in Subsections 5.1 and 5.2.

Subsequently, the bootstrap sample is utilized to compute the MISE for the unfolded distribution, where the unknown $\phi(x)$ is replaced by the average value of the unfolded distribution for the bootstrap sample. The regularization parameter $\hat{\lambda}$ is then defined as:
\begin{equation}
\hat{\lambda}= \argmin{\lambda_1,\ldots,\lambda_s} \,\, \mathrm{MISE}(\lambda).
\end{equation} 
By employing the bootstrap method and optimizing the MISE over a range of regularization parameters, the proposed approach aims to find a regularization parameter that provides a close-to-optimal estimation of the unfolded PDF.
\vspace*{1cm}
\section{Numeric example}
\subsection{Data simulation}
\begin{figure}

	\centering
	\begin{subfigure}{0.46\linewidth}
		\includegraphics[width=\linewidth]{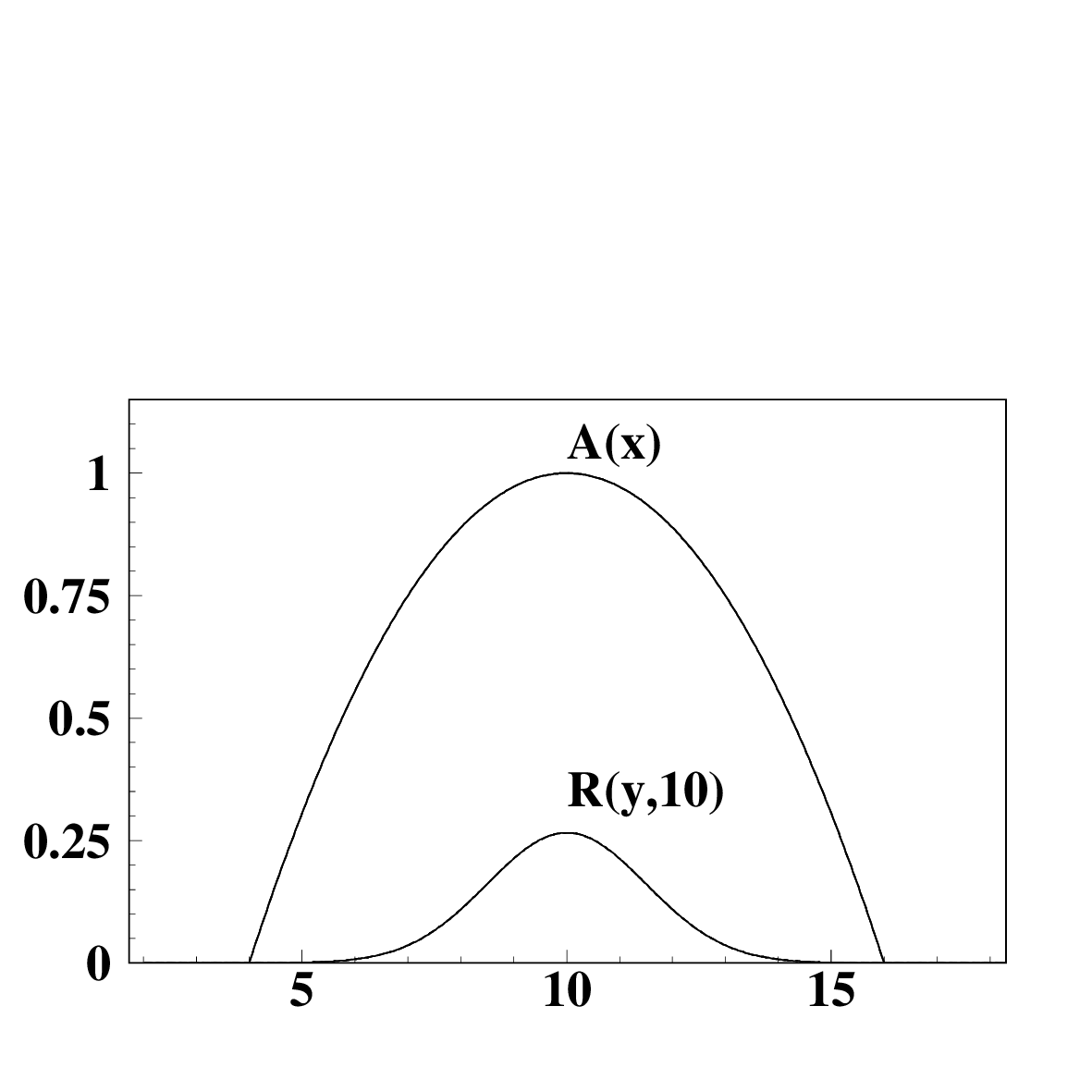}
	\end{subfigure}
   \hfill 
	\begin{subfigure}{0.46\linewidth}
\vspace{+1.5cm}
		\includegraphics[width=\linewidth]{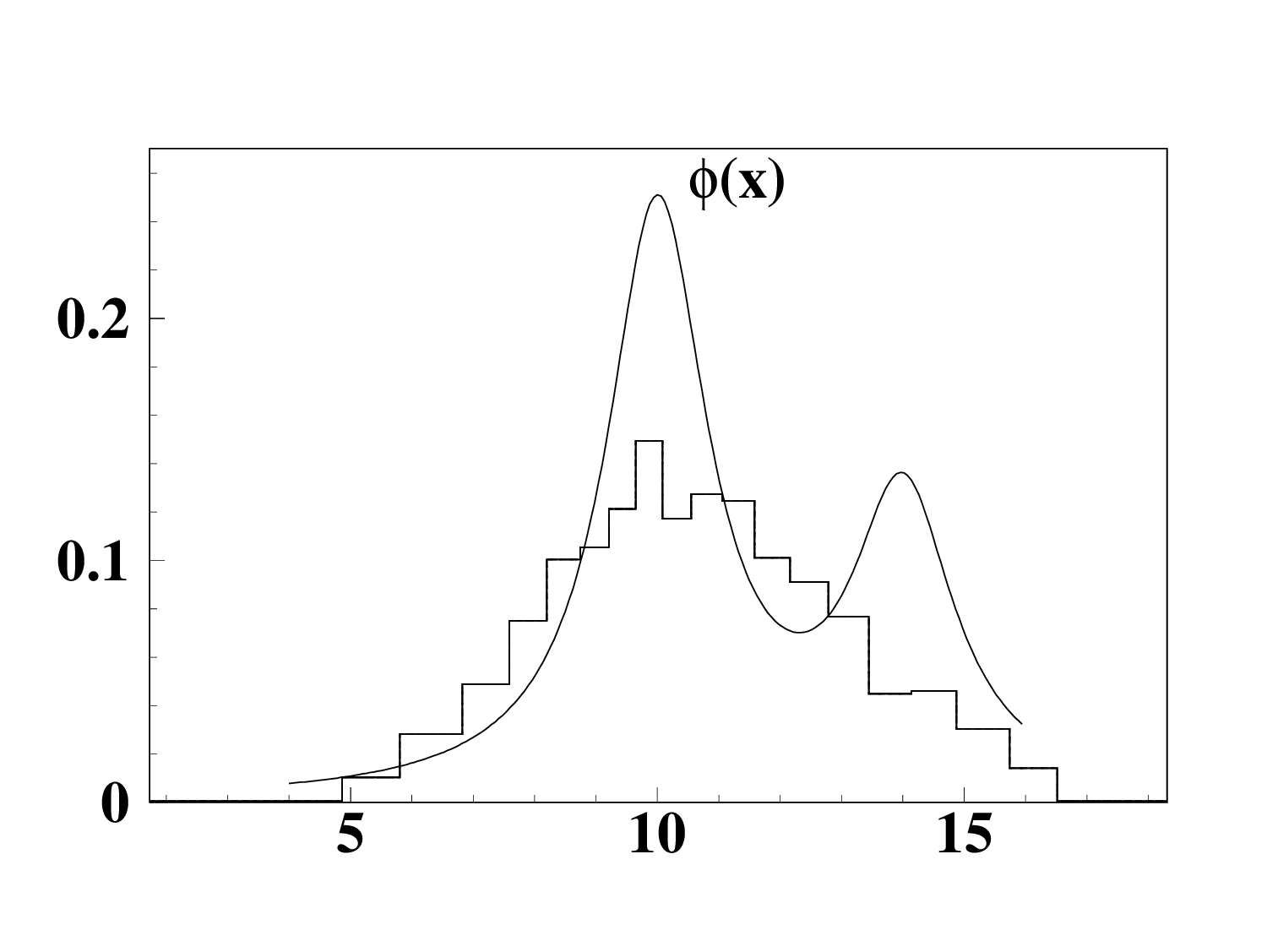}
	\end{subfigure}
	\caption{Acceptance $A(x)$\/ and resolution function $R(y,x)$
         for $x=10$ (left) and a histogram of the measured PDF based on a sample of $3000$ events generated
         for the true distribution $\phi(x)$ (right). The true distribution is shown  by the curve} 
	\label{fig:true}
\end{figure}

Following to \cite{zhig} let us assume a true PDF $\phi(x)$  that is described by a sum of two Breit-Wigner functions defined on interval $[4, 16]$:
\begin{equation}
 \phi(x) \propto 2\frac{1}{(x-10)^2+1} + \frac{1}{(x-14)^2+1}
\label{testform}
\end{equation}
from which the measured PDF $f(y)$  is obtained according to equation (\ref{fred})
with the acceptance function $A(x)$ (Figure~\ref{fig:true}):
\begin{equation}
A(x)=1-\frac{(x-10)^2}{36} \label{acc}
\end{equation}
and the resolution function describing Gaussian smearing (Figure~\ref{fig:true}):
\begin{equation}
R(y,x)=\frac{1}{\sqrt{2\pi}\sigma}\exp\left(-\frac{(y-x)^2}{2\sigma^2}\right), \, \sigma=1.5\; \label{res}  .
\end{equation}
The measured distribution obtained by
simulating  3000 events according to PDF $\phi(x)$  is  shown in Figure~\ref{fig:true}. A histogram with binning obtained by K-mean clustering 
method with a number of bins $l=20$ is used for that.
\newpage

\subsection{System identification}
To identify the measurement system, we simulate  it as described  by a sum of two Breit-Wigner functions defined as:
\begin{equation}
 \phi^s (x) \propto \frac{1.5^2}{(x-8)^2+1.5^2} + 2\frac{1.5^2}{(x-12)^2+1.5^2}  \label{rcs}  .
\end{equation}
From this, the reconstructed  PDF $f^s(y)$  is obtained according to equation (\ref{fred}) with the same acceptance  $A(x)$  equation  (\ref{acc}) and resolution function  $R(y,x)$  (\ref{res}).

\begin{figure}
\vspace*{-1cm}
	\centering
 	\begin{subfigure}{0.56\linewidth}
		\includegraphics[width=\linewidth]{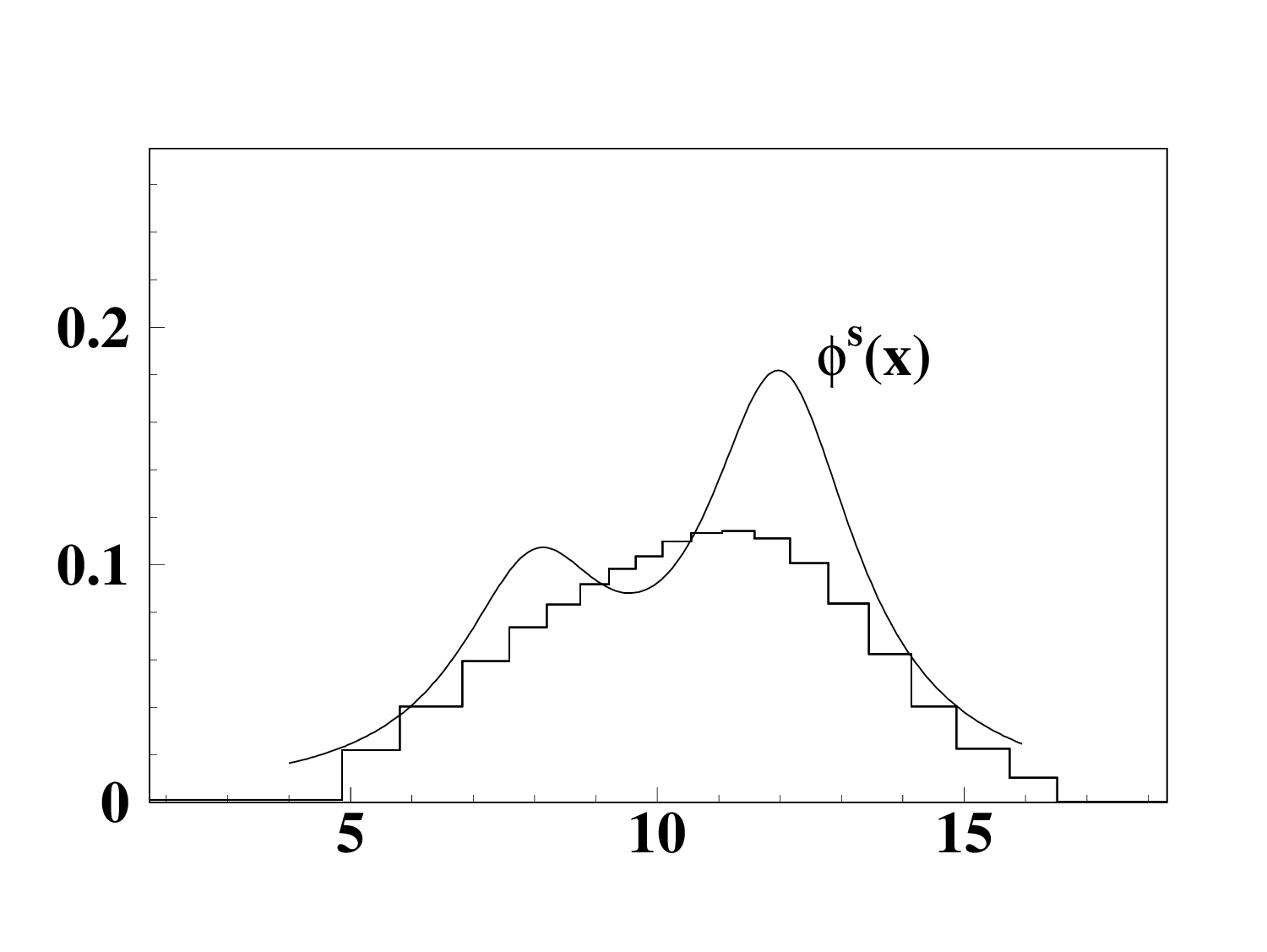}
	\end{subfigure}
   \hfill 
	\begin{subfigure}{0.4\linewidth}
		\includegraphics[width=\linewidth]{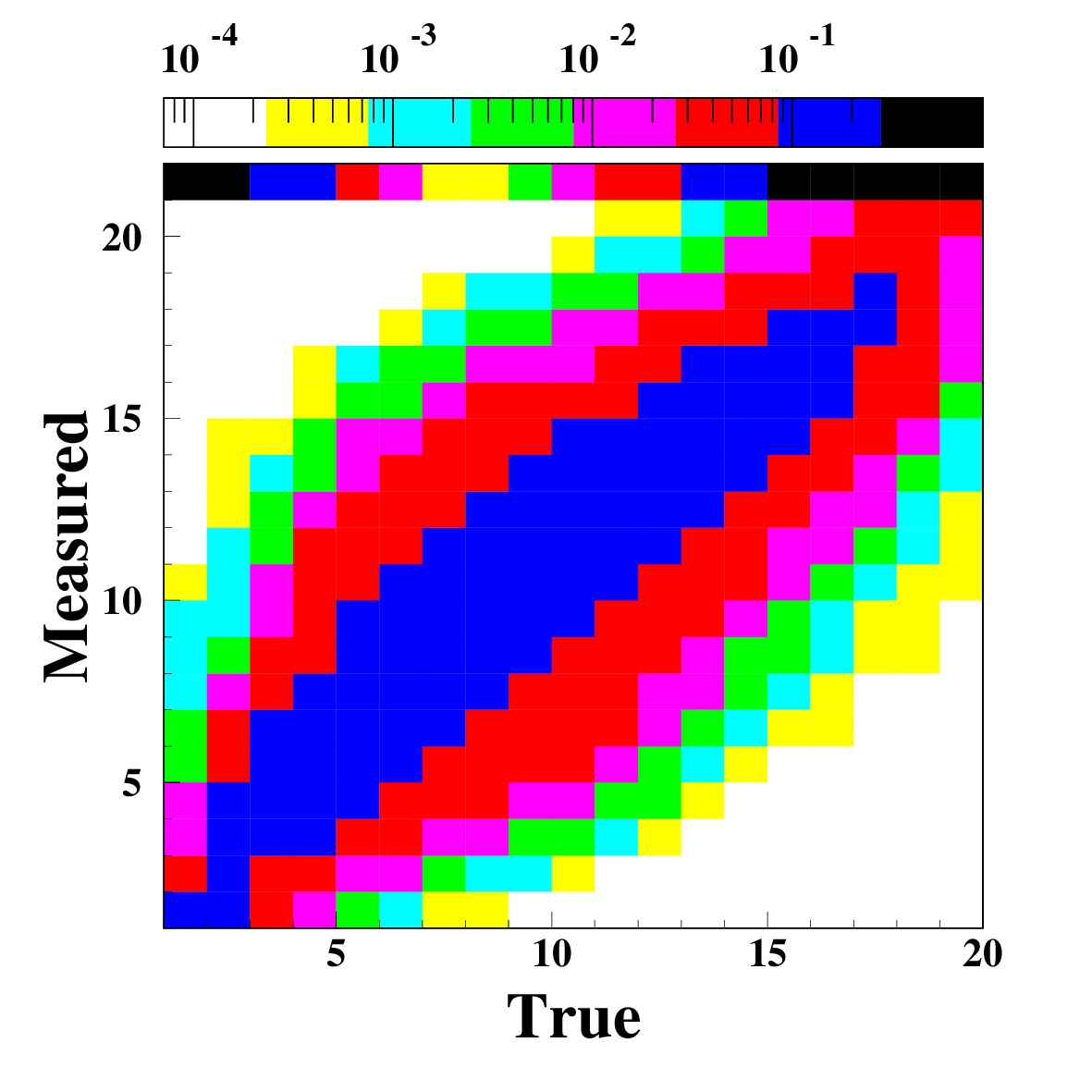}
	\end{subfigure}        
	\caption{A  histogram of the reconstructed PDF $f^s(x)$, the generated  distribution  $\phi^s(x)$ is shown  by the curve (left) and  matrix  $R$ calculated according algorithm described in Section 3 (right).}
	\label{fig:res}
\end{figure}

The reconstructed distribution obtained by simulating  of $10^6$  events according to PDF $\phi^s(x)$  is  shown in Figure~\ref{fig:res}.
The bin width of the impulse signal equal to $1.2$ with a total number of 50\% overlapping  impulse signals equal to 200. The matrix $R$ was calculated for 19 bins of unfolded distribution and 20 bins of measured distribution, with binning for both cases defined by K-mean clustering.  An  illustration of the matrix is presented on the same  Figure~\ref{fig:res}. The first row  presents matrix elements related to the efficiency of registration.  
\subsection{ Bootstrap resampling} 
For resampling the measured data by the parametric bootstrap  method,  a regularization parameter equal to $\lambda_{\textrm{bs}}=0.1$ was chosen according to criteria described in Section 6. 
 500 bootstrap samples were simulated using a multinomial random number generator \cite{cern} with probabilities 
\newpage
 \noindent 
$\hat{\theta}_i = \sum_{j=1}^m r_{ij} \hat{p}_j$ used as parameters, where $\hat{p}_i$ was obtained as result of unfolding with  $\lambda_{\textrm{bs}}=0.1$.
\subsection{Estimating the optimal regularization parameter}
To estimate the optimal regularization parameter, the Mean Integrated Squared Error (MISE) was studied using the method described in Section 6. The regularization parameter $\lambda$ was scanned over the interval $[0.15, 100]$. The result of the scan provided an optimal value of
$\hat{\lambda} = 6.95$. 

\begin{figure}
\vspace *{-2cm}
	\centering
 	\begin{subfigure}{0.46\linewidth}
		\includegraphics[width=\linewidth]{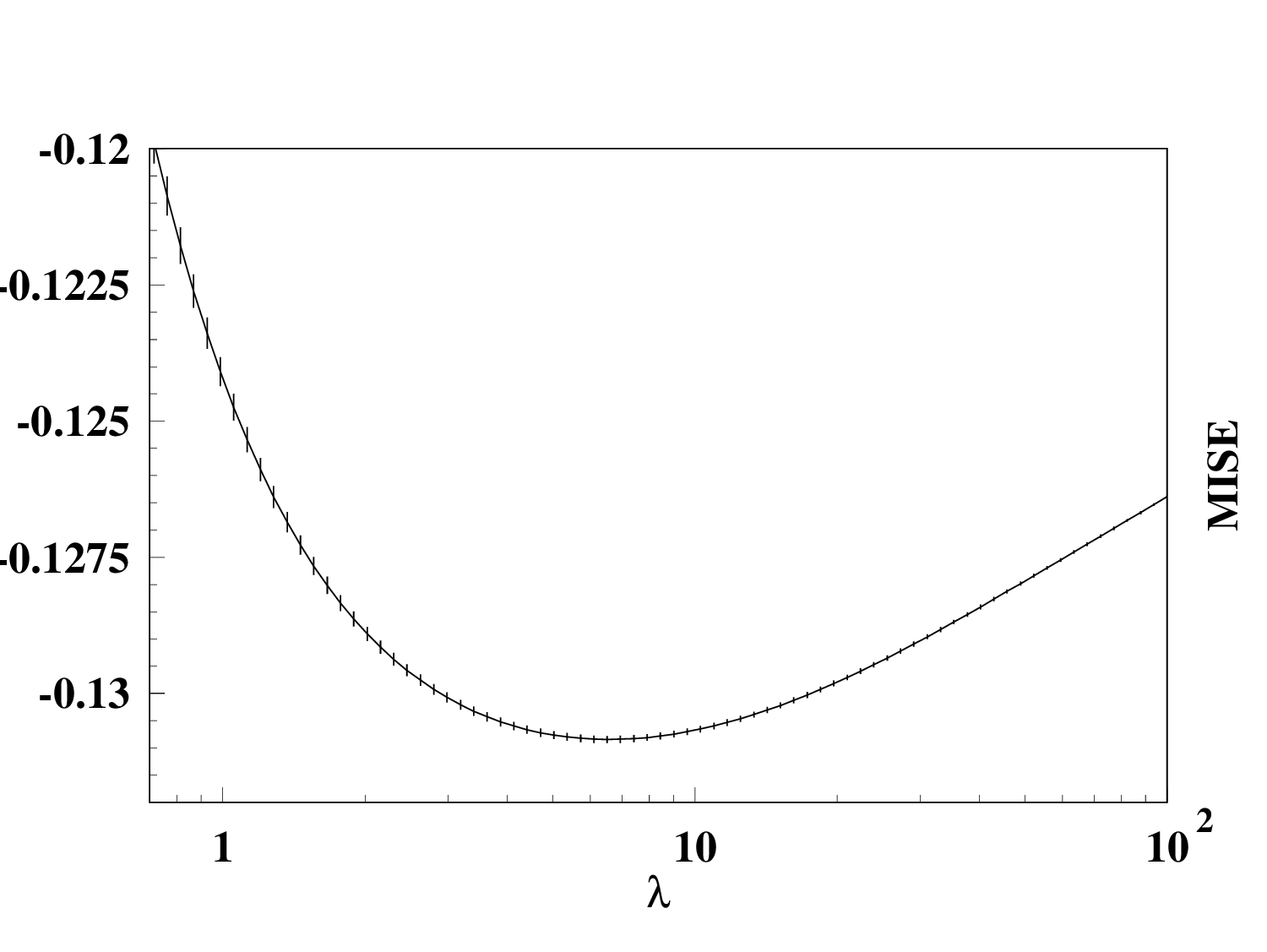}
	\end{subfigure}
   \hfill 
	\begin{subfigure}{0.46\linewidth}
		\includegraphics[width=\linewidth]{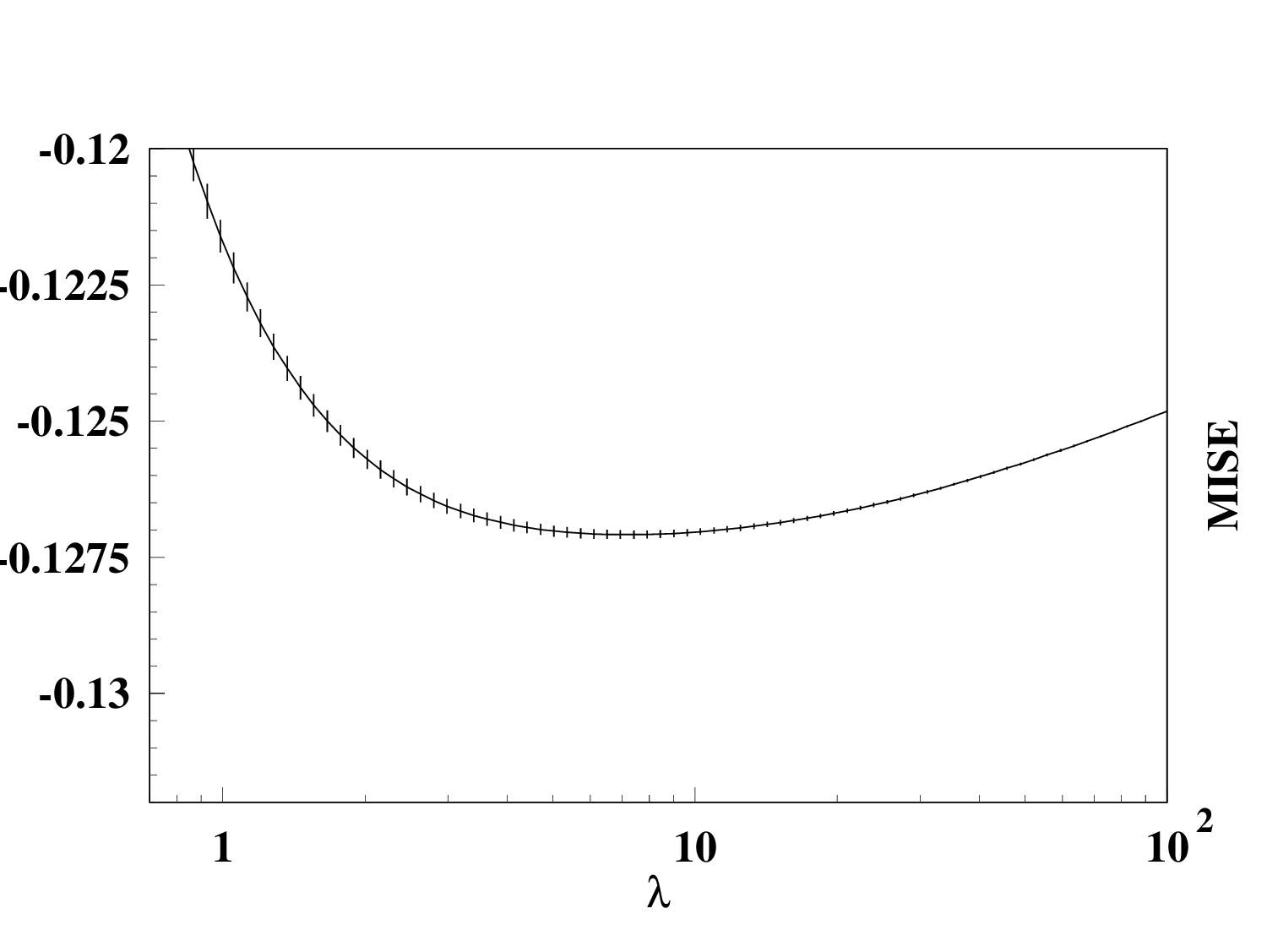}
	\end{subfigure}        
	\caption{MISE for simulated data (left) and for bootstrap data (right). 500 replications of the unfolded distribution are used for the MISE
estimation in both cases. The standard deviations of the estimated MISE are presented as error bars.
  }
	\label{fig:res1}
\end{figure}

For comparison  500 samples using true PDF $\phi(x)$ were simulated. Using the same procedure, the optimal value obtained is                                              
$\hat{\lambda} =6.52$.
Figure \ref {fig:res1} presented plots of MISE for both cases,  excluding the constant part of the expressions, which does not influence the result as noted in Section 5.  
\subsection {Data unfolding with estimation errors and correlation matrix}
Unfolding was performed for $\hat{\lambda}=6.95$ with the calculation of errors and the correlation matrix using bootstrap data. The same procedure was repeated for $\hat{\lambda}=6.52$, estimating the correlation matrix on simulated data (see Figures 4 and 5).  Additionally, the average unfolded distribution was produced using both simulated data and bootstrap data in both cases of 500 samples (see Figure 6). These calculations provide information about the bias for both cases.
\begin{figure}
	\centering
 	\begin{subfigure}{0.46\linewidth}
		\includegraphics[width=\linewidth]{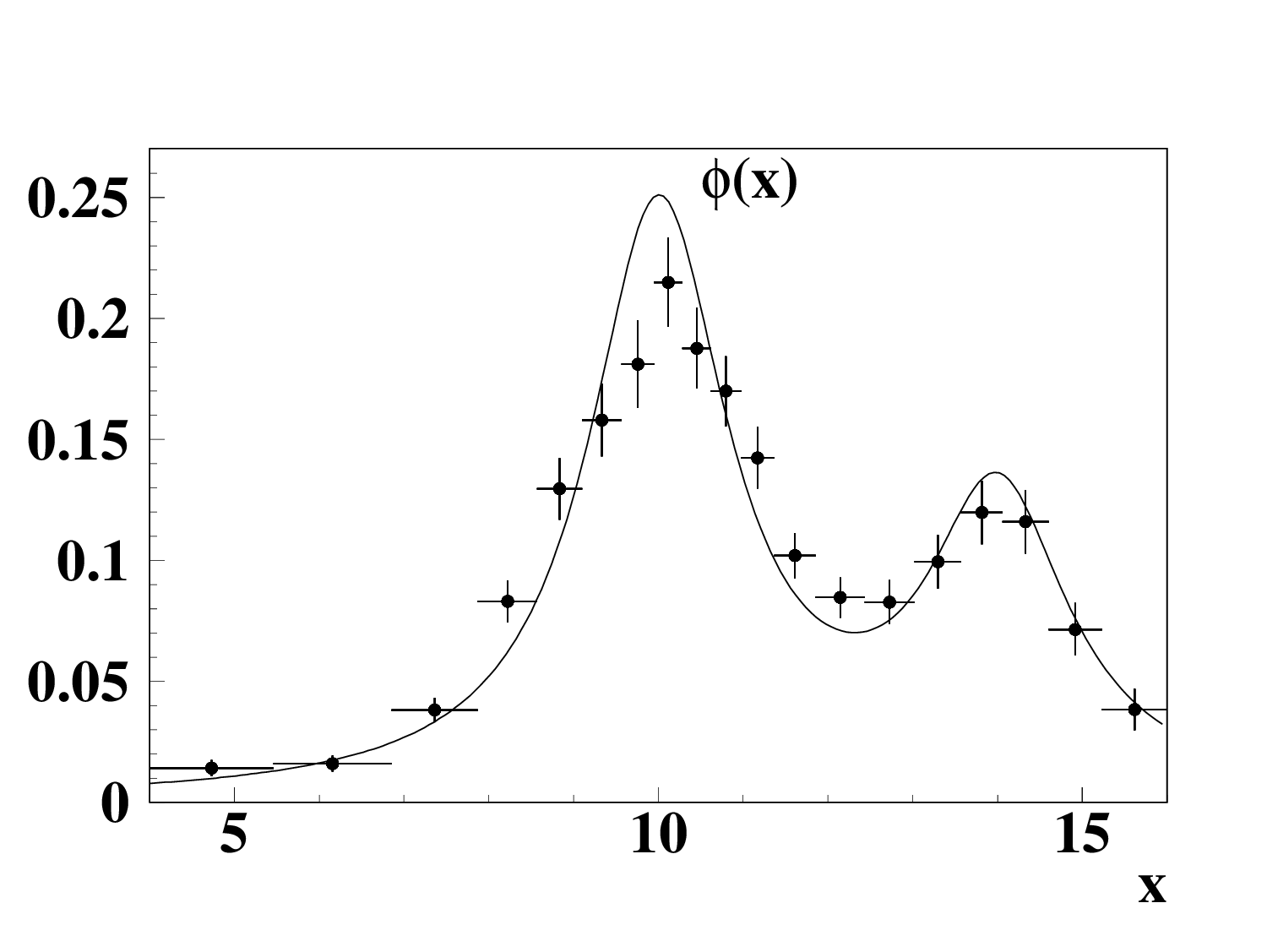}
	\end{subfigure}
   \hfill 
	\begin{subfigure}{0.46\linewidth}
		\includegraphics[width=\linewidth]{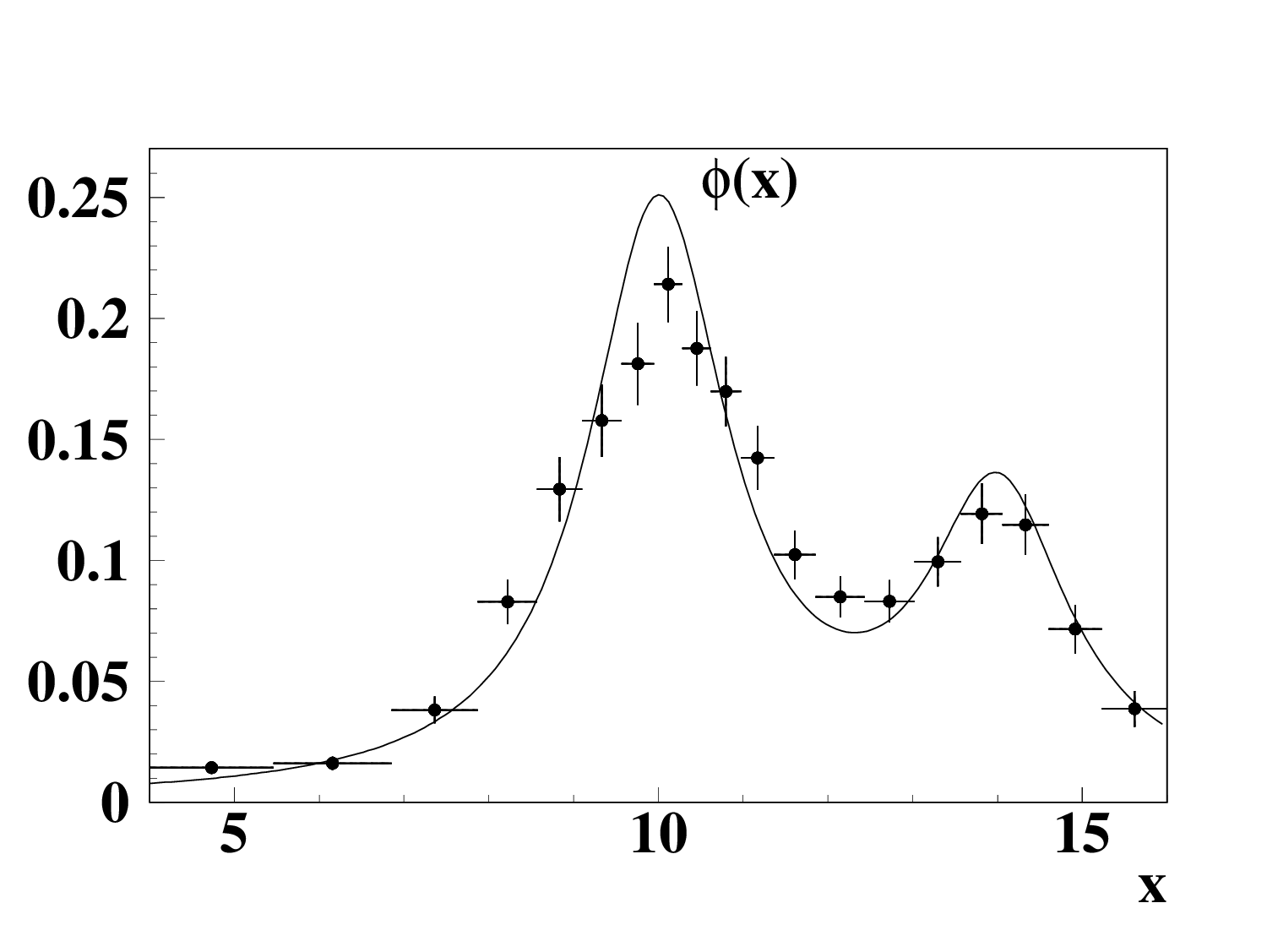}
	\end{subfigure}        
	\caption{Unfolded PDF for the  $\lambda=6.52$  (left)  and unfolded PDF for the  $\lambda=6.95$  (right). The true distribution  is shown  by the curve.}
	\label{fig:res2}
\end{figure}
\begin{figure}
	\centering
 	\begin{subfigure}{0.46\linewidth}
		\includegraphics[width=\linewidth]{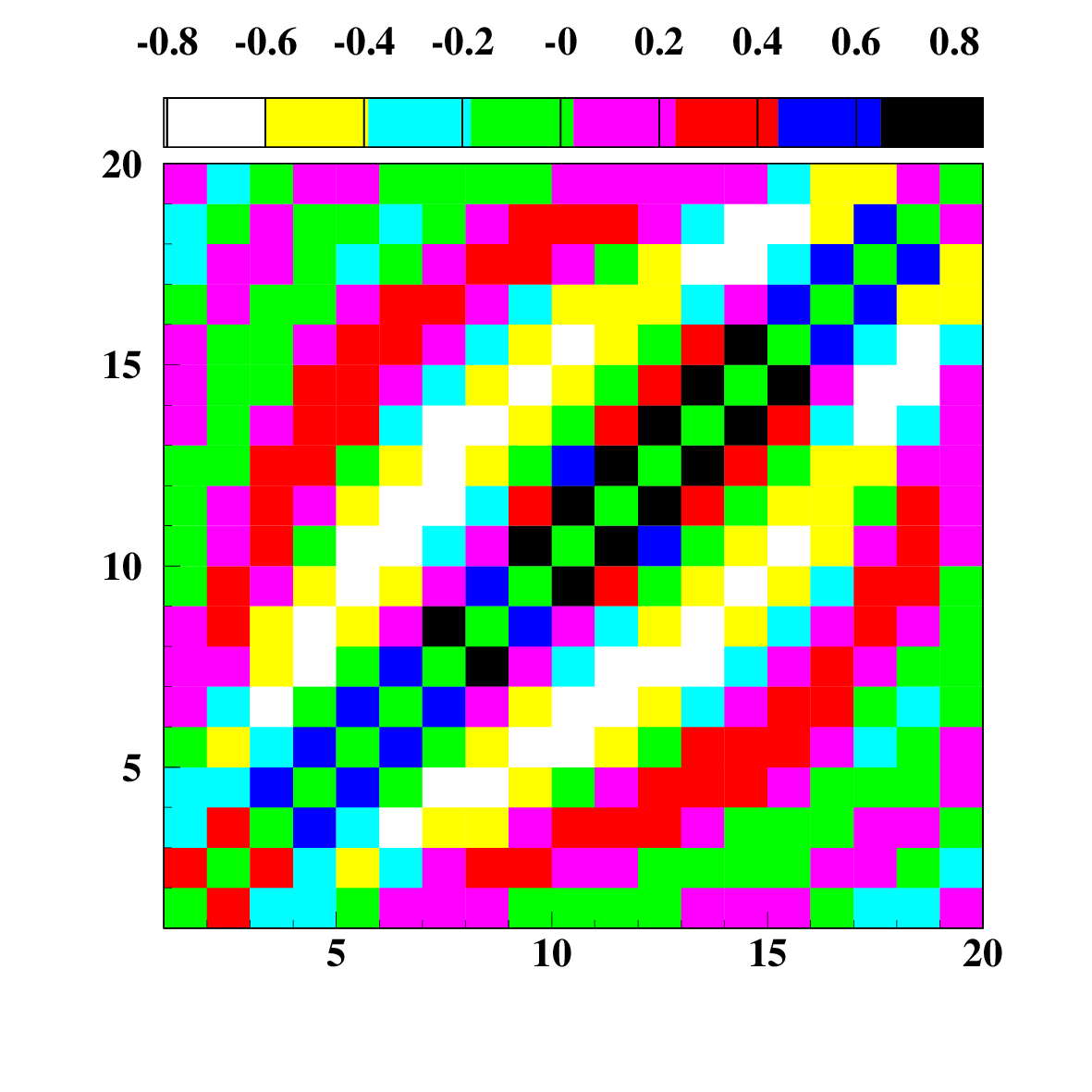}
	\end{subfigure}
   \hfill 
	\begin{subfigure}{0.46\linewidth}
		\includegraphics[width=\linewidth]{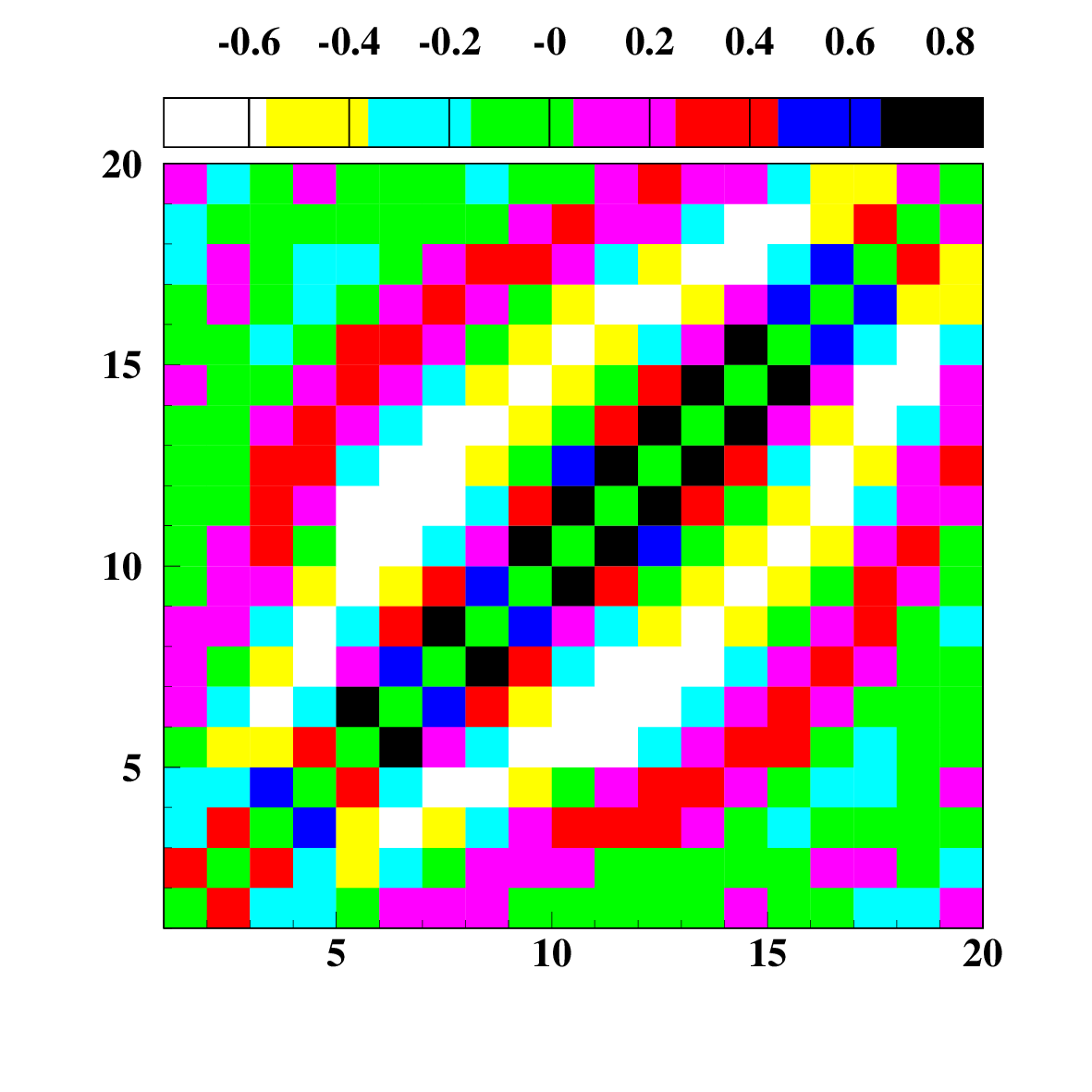}
	\end{subfigure}        
	\caption{Estimation of correlation matrix using simulated data (left) and  estimation of correlation matrix using bootstrap data (right).}
	\label{fig:res3}
\end{figure}
\newpage
\begin{figure}[ht]
	\centering
 	\begin{subfigure}{0.46\linewidth}
		\includegraphics[width=\linewidth]{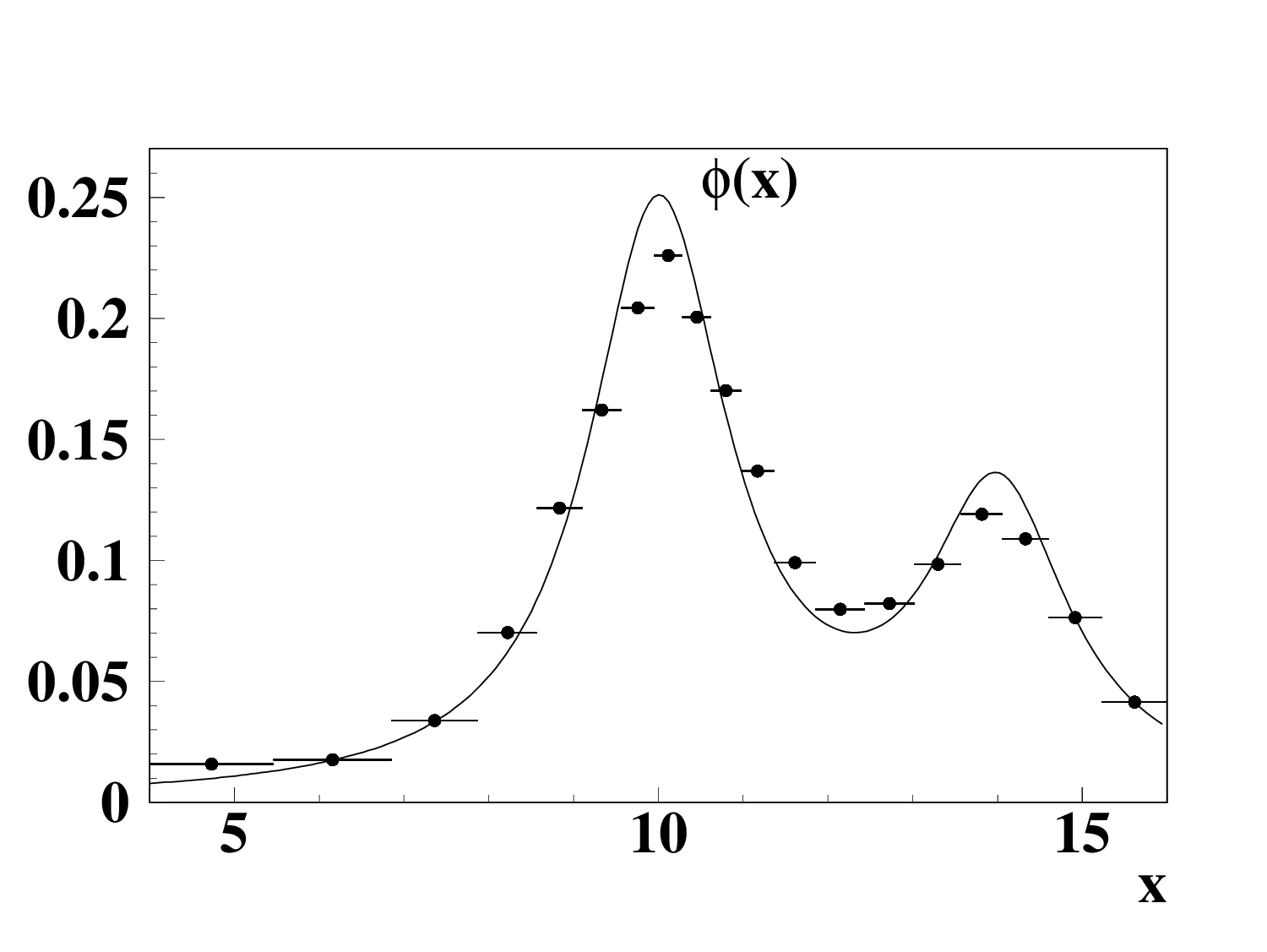}
	\end{subfigure}
   \hfill 
	\begin{subfigure}{0.46\linewidth}
		\includegraphics[width=\linewidth]{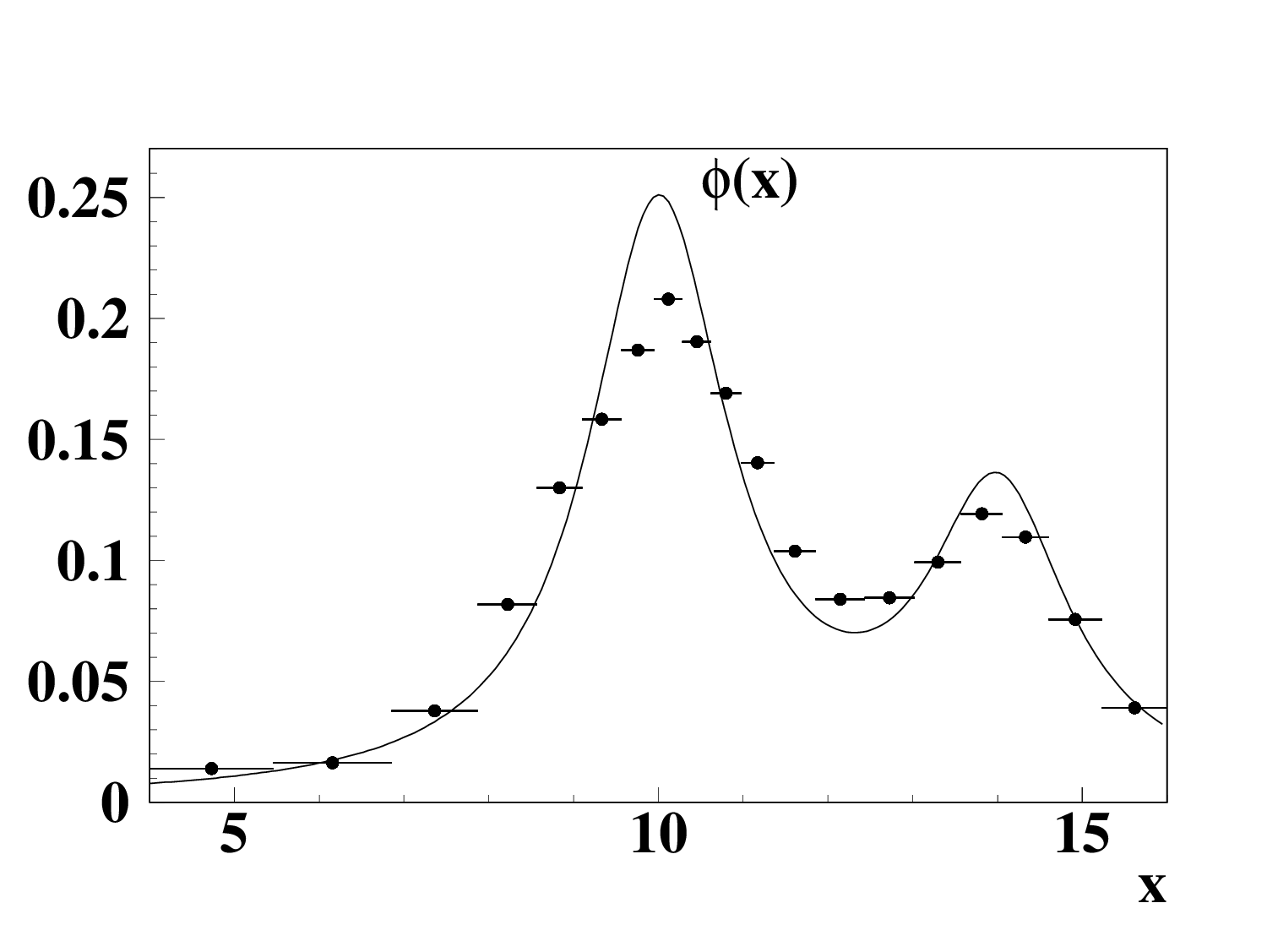}
	\end{subfigure}        
	\caption{Average unfolded distribution for simulated data (left) and  average unfolded distribution for bootstrap data (right).}
	\label{fig:res4}
\end{figure}
\section{Conclusions}
In this paper, unfolding is formulated as a Mean Integrated Square Error  optimization problem. We propose a bootstrap algorithm aimed at finding solutions close to the optimal solution. The bootstrap algorithm is used to estimate statistical errors and the correlation matrix of the solution as well. Additionally, we develop a methodology for creating a mathematical model of a measurement system using simulated data.This approach renders the experimental data, along with the measurement system model, reusable.   Furthermore, we devise a computational algorithm for nonlinear optimization, which is utilized to maximize the likelihood functional that includes a regularization component. Our discretization scheme permits the application of the method in multidimensional cases and allows for data acquired from different apparatuses measuring the same spectra to be analyzed.
\section {Acknowledgments}
 The author express their gratitude to Michael Schmelling (Max Planck Institute for Nuclear Physics) and Florin Maciuc (Horia Hulubei National Institute of Physics and Nuclear Engineering)  for their valuable insights and contributions related to unfolding techniques. Special thanks are also extended to Sigurdur Detlef J\'{o}nsson from Icelandic High Performance Computing Center  for his assistance in overcoming the computationally intensive aspects of this work. Additionally, the author would like to express his gratitude to the anonymous reviewer for their valuable feedback and constructive comments, which have significantly improved the quality of this manuscript.
\newpage
\bibliographystyle{elsart-num}
\bibliography{project}

\begin{thebibliography}{10}

\bibitem{review}
L.~Brenner, R.~Balasubramanian, C.~Burgard, W.~Verkerke, G.~Cowan,
  P.~Verschuuren, and V.~Croft.
\newblock Comparison of unfolding methods using {RooFitUnfold}.
\newblock {\em International Journal of Modern Physics A}, 35(24):2050145,
  2020.

\bibitem{silver}
B.~W. Silverman.
\newblock {\em Density estimation for statistics and data analysis}.
\newblock Chapman and Hall, 1986.

\bibitem{fridman}
T.~Hastle, R.~Tibshirani, and J.~Friedman.
\newblock {\em The Elements of Statistical Learning: Data Mining, Inference,
  and Prediction}.
\newblock Springer, 2017.

\bibitem{taras}
M.~Z. Tarasko.
\newblock On the method for solution of the liner system with stochastic
  matrixes, {{Obninsk}, {PEI-156}}.
\newblock 1969.

\bibitem{Richardson}
W.~H. Richardson.
\newblock Bayesian-based iterative method of image restoration.
\newblock {\em J. Opt. Soc. Am.}, 62(1):55--59, 1972.

\bibitem{Lucy}
L.~B. {Lucy}.
\newblock {An iterative technique for the rectification of observed
  distributions}.
\newblock {\em The Astronomical Journal}, 79:745, 1974.

\bibitem{kart}
A.~Hocker and V.~Kartvelishvili.
\newblock {SVD} approach to data unfolding.
\newblock {\em Nucl. Instrum. Meth. A}, 372:489, 1996.

\bibitem{Tikhonov77}
A.~N. Tikhonov and V.~Y. Arsenin.
\newblock {\em Solutions of Ill-posed problems}.
\newblock W.H.~Winston, 1977.

\bibitem{bintobin}
V.~B. Anykeyev, A.~A. Spiridonov, and V.~P. Zhigunov.
\newblock Correcting factors method as an unfolding technique.
\newblock {\em Nucl. Instrum. Meth. A}, 322(2):280, 1992.

\bibitem{cluster}
P.~N. Tan, M.~Steinbach, A.~Karpatne, and V.~Kumar.
\newblock {\em Introduction to Data Mining}.
\newblock Pearson, 2018.

\bibitem{ident}
D.~Graupe.
\newblock {\em Identification of system}.
\newblock Krieger and Huntington,, 1976.

\bibitem{voronoi}
G.~Voronoi.
\newblock Nouvelles applications des parametres continus a la theorie des
  formes quadratiques.
\newblock {\em J. reine angew. Math.}, 133:97, 1907.

\bibitem{kuus2}
M.~Stanley, P.~Patil, and M.~Kuusela.
\newblock Uncertainty quantification for wide-bin unfolding: one-at-a-time
  strict bounds and prior-optimized confidence intervals.
\newblock 2022.
\newblock arXiv:2111.01091.

\bibitem{mateev}
P.~Mateev.
\newblock On the entropy of a multinomial distribution.
\newblock {\em Teor. Veroyatnost. i Primenen.}, 23(1):196, 1978.

\bibitem{multi}
J.~Cichon and Z.~Golebiewski.
\newblock On {B}ernoulli sums and {B}erstein polynomials.
\newblock {\em DMTCS proc. AQ}, page 179, 2012.

\bibitem{baker}
S.~Baker and R.~D. Cousins.
\newblock Clarification of the use chi-square and likelihood functions in fits
  to histograms.
\newblock {\em Nucl. Instrum. Meth.}, 221:437, 1984.

\bibitem{nielsen}
F.~Nielsen and V.~Garsia.
\newblock Statistical exponential families: A digest with flash cards.
\newblock 2011.
\newblock arXiv:0911.4863.

\bibitem{boot}
B.~Efron.
\newblock Bootstrap methods: another look at the jackknife.
\newblock {\em Annals of Statistics}, 7:126, 1979.

\bibitem{cern}
{CERN} {P}rogram {L}ibrary (v138).
\newblock http://cernlib.web.cern.ch/cernlib.

\bibitem{zhig}
V.~P. Zhigunov.
\newblock Improvement of resolution function as an inverse problem.
\newblock {\em Nucl. Instrum. Meth. A}, 322(1-2):183, 1983.

\end{thebibliography}
\end{document}